\documentclass[12pt]{article}

\usepackage{newtxtext,newtxmath}

\usepackage{graphicx}

\usepackage[letterpaper,margin=1in]{geometry}

\renewenvironment{abstract}
	{\quotation}
	{\endquotation}

\date{}

\makeatletter
\renewcommand{\fnum@figure}{\textbf{Figure \thefigure}}
\renewcommand{\fnum@table}{\textbf{Table \thetable}}
\makeatother

\usepackage{scicite}

\usepackage{url}

\usepackage[colorlinks=true, linkcolor=blue, urlcolor=blue]{hyperref}

\renewcommand{\deg}{^{\circ}}

\newcommand*\aap{Astron.~Astrophys.}%
\newcommand*\aapr{Astron.~Astrophys.~Rev.}%
\newcommand*\apj{Astrophys.~J.}%

\newcommand*\apss{Astrophys.~Space~Sci.}%
\newcommand*\grl{Geophys.~Res.~Lett.}%
\newcommand*\jgr{J.~Geophys.~Res.}%
\newcommand*\nat{Nature}%
\newcommand*\solphys{Sol.~Phys.}%
\newcommand*\ssr{Space~Sci.~Rev.}%

\newcommand{\erass}{eRASS\,}

\def\scititle{Determination of the Solar System contribution to the soft X--ray sky}

\title{\bfseries \boldmath \scititle}

\author{
	K. Dennerl$^{1\ast}$,
	G. Ponti$^{2,1}$,
	X. Zheng$^1$,\and
	M. J. Freyberg$^1$,
	S. Friedrich$^1$,
	Th. M{\"u}ller$^1$,
	M. C. H. Yeung$^1$\and
  \small$^{1}$Max-Planck-Institut f\"ur extraterrestrische Physik, Garching, Germany.\and
  \small$^{2}$Osservatorio Astronomico di Brera - Istituto Nazionale di Astrofisica (INAF), Merate, Italy.\and
	\small$^\ast$Corresponding author. Email: kod@mpe.mpg.de
}

\begin{document} 
    
\maketitle

\vspace*{-5mm}\noindent
\centerline{\vbox{\hsize=0.58\textwidth\noindent
This is the author's version of the work. It is posted here by permission of the AAAS for personal use, not for redistribution. The definitive version was published on 16 April 2026 in Science Vol 392, Issue 6795, pp.~285-288, \href{https://www.science.org/doi/10.1126/science.adt9147}{DOI: 10.1126/science.adt9147}.
}}


\begin{abstract} \bfseries \boldmath

Solar wind charged particles interact with diffuse gas within the heliosphere, producing soft X--rays. This solar wind charge exchange (SWCX) process produces foreground emission that complicates interpretation of X--ray observations. In this work, we analyze X--ray observations of the western Galactic hemisphere by the Extended Roentgen Survey with an Imaging Telescope Array (eROSITA) instrument on the Spectrum-Roentgen-Gamma (SRG) spacecraft. These data avoid contamination by Earth's geocorona and are derived from four surveys of the full sky, including during the minimum of the Sun's activity cycle. We determine the SWCX contribution and subtract it from the survey, providing a less contaminated view of the diffuse soft X--ray sky. We also demonstrate that X--rays can be used to map the flow of interstellar matter through the Solar System.

\end{abstract}


\noindent
Astronomical surveys in soft X--rays [photon energy $E\lesssim 1\mbox{ keV}$] \cite{sno97a,zhe24a} are contaminated by a diffuse, inhomogeneous, and time-variable foreground emission that originates in the Solar System. This foreground emission results from interactions between the solar wind and diffuse gas within the heliosphere. In this process, known as solar wind charge exchange (SWCX) \cite{cra97a}, highly charged ions from the solar wind encounter atoms or molecules that are less ionized or neutral; electron transfer then leaves the ions in an excited state, which de-excites through processes that include the emission of soft X--ray photons. Owing to the high cross section for SWCX \cite{cra97a}, even very tenuous gas emits substantial X--rays. Such gas is present around comets \cite{lis96d}, the upper atmosphere of Earth (its geocorona) \cite{kam17a} and within the entire Solar System (the heliosphere) \cite{que14a}. The geocoronal and heliospheric SWCX foregrounds \cite{den97a,cox98a,fre98a} are a major contamination of the soft diffuse X--ray sky. Previous quantitative estimates of the SWCX contribution have provided only a partial correction, with large uncertainties \cite{cra00a,wel09b,upr16a,sib18a,kun19a}.

The extended Roentgen Survey with an Imaging Telescope Array (eROSITA) instrument on the Spectrum-Roentgen-Gamma (SRG) spacecraft \cite{pre21a}, a soft X--ray space telescope, is located about 1.5~million kilometers from Earth (Fig.~\ref{fig:orbit}), well outside the geocorona. It has mapped the full sky four times, beginning in December 2019 when the Sun was at the minimum of its 11-year activity cycle \cite{mer24a}. We analyzed the eROSITA observations of the western Galactic hemisphere \cite{pre21a} in the energy range of 0.27 to 0.69~keV, where eROSITA is most sensitive to SWCX \cite{methods}. 

\subsection*{Determining the heliospheric emission component}

Because the heavy ion content of the solar wind is highly variable \cite{ste10a}, heliospheric SWCX emission exhibits temporal, spatial, and spectral variability. We used the western Galactic hemisphere data from four eROSITA All Sky Surveys (\erass1 to \erass4)\cite{methods} to separate the variable component. Any given location on the sky was typically observed $\ge$ 5~times per day during four observing intervals, each separated by 6~months, with the duration of each interval ranging from 1~day at the ecliptic plane to 6 months at the ecliptic poles. Many unrelated  X--ray sources exhibit temporal variability on similar timescales, but these sources appear essentially point-like in the eROSITA data. We therefore assumed that there is negligible variability of the diffuse X--ray sky (on scales of square degrees) from sources beyond the heliosphere. Because there is also negligible absorption inside the heliosphere \cite{que14a,note:heliosphere}, the variability results from additive positive flux components. We minimized their contribution by selecting the survey scan with the lowest flux in each location within the western Galactic hemisphere, and then we combined those to produce a dark sky map \cite{methods}.

By construction, the dark sky map contains less flux than any individual eRASS map. However, residual heliospheric emission cannot be entirely excluded because it is possible that each of the four surveys contained nonzero SWCX flux. The presence of SWCX with a variability on timescales of $<2\mbox{ years}$ (the time between \erass1 and \erass4) is unlikely because that would have produced some modulation in the dark sky map, preferentially as a function of ecliptic longitude (see the supplementary text), which we do not observe. However, the dark sky map could contain a component that is constant or changing on a timescale of several years or more. We seek to constrain and quantify any residual flux by subtracting the dark sky map from the individual eRASS maps and then determining the remaining flux as a function of ecliptic latitude (Fig.\,\ref{fig:profiles}A, discussed below). We find a continuous increase of flux starting from \erass3, which is most pronounced at low latitudes. 

This is in qualitative agreement with our expectations. The solar wind consists predominantly of a slow and a fast component. From solar minimum to maximum, the slow component expands from low to high latitudes; the other regions are occupied by the faster solar wind \cite{mcc08b}. The slow solar wind is more ionized than the fast solar wind: the O$^{7+}$/O$^{6+}$ and C$^{6+}$/C$^{5+}$ ratios are an order of magnitude higher [ \cite{kou24a}, their figure 3 ], so the slow solar wind generates more SWCX emission in the eROSITA energy band (supplementary text). We therefore ascribe the trend from \erass2 to \erass4 in Fig.\,\ref{fig:profiles}A to latitudinal expansion of the slow solar wind. However, for \erass1, there is lower flux at $<20\deg$ latitude compared with at higher latitudes, which we did not expect. The simplest explanation is the presence of residual SWCX emission in the dark sky map at these latitudes, which has caused too much flux to be subtracted. Figure \ref{fig:profiles}A allows us to quantify the residual SWCX emission to $\sim1\mbox{ line unit}$ (LU, defined as photons per square centimeter per second per steradian) \cite{methods}. This is $\sim6\%$ of the flux observed from the darkest sky regions in the western Galactic hemisphere and close to the eROSITA sensitivity limit (supplementary text). We therefore conclude that, even in the most contaminated regions, $\gtrsim94\mbox{\%}$ of the flux in the dark sky map is from beyond the Solar System.

\subsection*{Properties of the heliospheric emission}

After confirming that the dark sky map is largely free of heliospheric emission, we isolated the SWCX component by subtracting that map from the individual eRASS maps (Fig.\,\ref{fig:heliosphere}\,A-D). The resulting images (Fig.\,\ref{fig:heliosphere}\,E-H) show how the long--term brightening and latitudinal expansion (discussed above) appears when spatially resolved, after smoothing along the scanning direction (Fig.\,\ref{fig:heliosphere}\,I-L). This is consistent with in situ solar wind measurements [\cite{mcc08b}, their figure~1~], which imply that an X--ray dark region is produced by the less ionized fast solar wind at high latitudes, which shrinks as the solar cycle progresses. Figure~\ref{fig:heliosphere} also shows that the heliospheric emission exhibits short--term temporal flux variations in addition to the long--term evolution. Because the eROSITA energy bands are mainly sensitive to emission from carbon, nitrogen, and oxygen ions (supplementary text), we ascribe the variations in flux to abundance variations of carbon, nitrogen, and oxygen in the solar wind.

\subsection*{Localization of the emitting regions}

While eROSITA was scanning the sky, the SRG spacecraft moved along its orbit around the Sun. We used this to constrain the locations of the gas responsible for the heliospheric emission. For this analysis, we approximated the solar wind by a constant flow \cite{methods} and assumed that the emitting regions are stationary and have constant luminosity, so the observed variability is a result of the changing viewpoint around the orbit. The distance to the emitting region was then be determined from the parallax effect (fig.\,\ref{fig:loc_demo}). We extracted the longitudinal profile of the heliospheric emission by averaging the \erass3 and \erass4 data (where SWCX is highest) over $\pm20\deg$ in ecliptic latitude (Fig.~\ref{fig:loc_result}\,B-C), and we then searched for a stationary emissivity pattern that reproduces the observed flux profiles. We tested a superposition of 72 potential emission locations up to 4.3~astronomical unit (au) radius from the Sun, all on the ecliptic plane. Each potential location was assigned a trial emissivity (required to be positive), and then we computed the resulting emission map using bicubic spline interpolation in polar coordinates and derived the longitudinal profile. By comparing the trial profiles to the observations, we determined the best-fitting 72 emissivity parameters \cite{methods}.

The best-fitting emissivity map (Fig.\,\ref{fig:loc_result}A) shows a pattern that is almost mirror symmetric. The symmetry axis is close to the direction of motion of the Sun through the local interstellar medium (LISM), which is toward heliocentric ecliptic longitude $254.7\deg$ and latitude $+5.2\deg$, with relative velocity $26\mbox{ km s}^{-1}$ \cite{lal05b}. The Sun interacts with the flow of the LISM through gravitation, electromagnetic radiation, and the solar wind. Helium atoms in the LISM are deflected by the gravitational field, which causes them to move along ballistic trajectories around the Sun, forming a distribution known as the helium focusing cone \cite{wel74a,lal04a}. Hydrogen atoms are more susceptible to ionization by radiation and charge exchange with solar wind protons; once ionized, they are swept up by the solar wind, forming a hydrogen cavity around the Sun \cite{ber71a,tho71a}.

The brightest emission region in Fig.~\ref{fig:loc_result} is close to the sunward part of the helium focusing cone, where we expect the emissivity to be highest owing to the expansion of the solar wind. There is also emission in the opposite direction, close to the hydrogen maximum emissivity region (MER) \cite{que14a}, beyond the edge of the hydrogen cavity. These regions are already known to backscatter emission in the hydrogen Lyman $\alpha$ (Ly-${\alpha}$) line, and previous work has identified possible SWCX X--ray enhancements by the He focusing cone \cite{kou09a,gal14a,upr16a}.

By rotating the locations of emission centers (in step of $1\deg$) around the Sun and repeating the fitting process (movie~S1), we determined that the brightest emission is at longitude $61.0\pm4.2\deg$ and the opposing emission is symmetrical about $246.2\pm5.8\deg$. These values differ by $13.7\deg$ and $6.0\deg$ from previous measurements of the He and H flow \cite{lal05b}. We interpret these offsets as a result of the SWCX emission being dominated by spiral density enhancements (see below), making the SWCX emission inhomogeneous (movie~S2).

\subsection*{Comparisons with simulations}

We use the LISM distribution and solar wind measurements to construct a simulation of the expected heliospheric emission \cite{methods}. We do not expect this to exactly match the observations for several reasons:
({\bf i})~The entry of the LISM into the Solar System depends on solar activity, which changes over the solar cycle \cite{lal04b}, but we assumed a static approximation of the hydrogen and helium distributions [\cite{rin21a}, their figure 1].
({\bf ii})~The input solar wind data were extracted from measurements by spacecraft near Earth
\cite{kin05a}. We then extrapolated them to the rest of the inner Solar System \cite{methods}.
({\bf iii})~The SWCX X--ray emission is caused by highly ionized heavy atoms, but the input solar wind data only measured protons. Previous measurements have shown that these are only weakly correlated [\cite{boc07a}, their figure 3]. Even if heavy ion measurements were available, the relevant charge exchange cross sections are highly uncertain.
({\bf iv})~Outside the ecliptic plane, even proton flux measurements were unavailable.

To address issue (iii), we introduced an empirical SWCX efficiency, which relates the product of proton flux and LISM density to SWCX volume emissivity. By combining the eROSITA measurements with our simulation \cite{methods}, we constrained the SWCX efficiency using an iterative process. Beginning with assumed values, we computed volume emissivities, integrated them along the line of sight, derived surface brightness profiles, compared them with the measurements, then adjusted the SWCX efficiencies until we were satisfied with the agreement between the two  (Fig.\,\ref{fig:profiles}). We found that the SWCX efficiency needed to be latitude and time dependent, initially rising steeply at the beginning of \erass3 at latitudes $<20\deg$, then expanding to $\sim40\deg$ at the end of \erass4 (fig.\,\ref{fig:pleff}). These values are heliocentric latitudes because this method also reconstructed  the heliocentric latitudinal SWCX efficiency profiles from the observed spacecraft-centric latitudinal surface brightness profiles \cite{methods}.

Figure~\ref{fig:simulation} shows the simulation output for a single day (21 September 2021) -- a map of the collision rate density per unit cross section (Fig.\,\ref{fig:simulation}\,A) and the resulting SWCX surface brightness sky map in the eROSITA energy band (Fig.\,\ref{fig:simulation}\,B).  Movies~S3 and S4 show an animation of the simulation over the full \erass1\,--\,\erass4 period. Movie~S5 shows how the collision rate density map changes when averaged over an extended time period.

We draw several conclusions from Fig.\,\ref{fig:simulation} and movies~S3\,--\,S5:
({\bf i})~Most heliospheric SWCX X--ray emission occurs in the inner Solar System \cite{note1}
({\bf ii})~When observations are taken approximately perpendicular to the direction of the Sun (as is typical in X--ray observations), there is little SWCX enhancement caused by the helium focusing cone.
({\bf iii})~At any given time, the product of solar wind flux and interstellar matter density is dominated by spiral density enhancements. The helium focusing cone becomes the dominant structure only when averaged over more than a few days.
({\bf iv})~The solar cycle dependence of long--term heliospheric SWCX emission is particularly pronounced a low ecliptic latitudes.
({\bf v})~Heliospheric SWCX variability at low ecliptic latitudes depends on the viewing direction. It is reduced when looking forward along the direction of Earth's movement, but increased when looking backward along the spiral structures.
({\bf vi})~In the backward direction, heliospheric SWCX occasionally produces a pattern of tightly localized enhancements, which can cause variability on timescales of hours when it sweeps across the sky. Previous studies of SWCX variability, which were also affected by geocoronal emission, have often attributed short--term changes to Earth's geocorona and long--term changes to the heliosphere \cite{kun19a}. However, because of the absence of geocoronal emission in the eROSITA data, we conclude that that heliospheric SWCX is also capable of generating short--term variability.

\clearpage


\begin{figure}
\vspace*{-8mm}
\centerline{\includegraphics[width=0.8\textwidth]{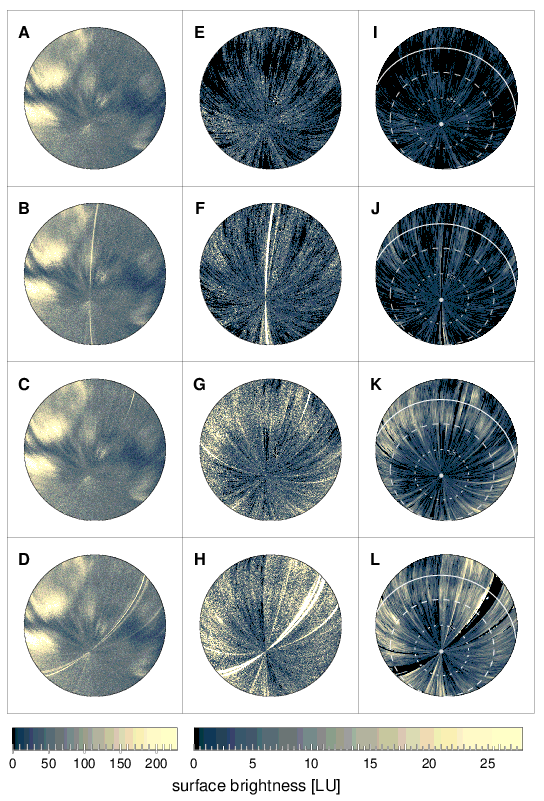}}
\caption{\textbf{Extracting the heliospheric SWCX emission.}
All panels show 0.27\,--\,0.69~keV maps of the western Galactic hemisphere 
in Lambert zenithal equal area projection, with the Galactic Center at the left edge and the Galactic plane running horizontally, at different processing steps. {\bf (A-D)} Maps obtained from \erass1 to \erass4 (respectively), used to derive the dark sky map. {\bf (E-H)} Corresponding maps after subtracting the dark sky map, on a different color scale. {\bf (I-L)} Corresponding maps after masking out periods of enhanced particle background and smoothing along the scan direction. Ecliptic latitudes $0^{\circ}$, $-30^{\circ}$ ,$-60^{\circ}$, and $-90^{\circ}$ are marked by solid, dashed, and dotted lines and a dot, respectively.}
\label{fig:heliosphere}
\end{figure}


\begin{figure}
\vspace*{-8mm}
\centerline{\includegraphics[width=0.75\textwidth]{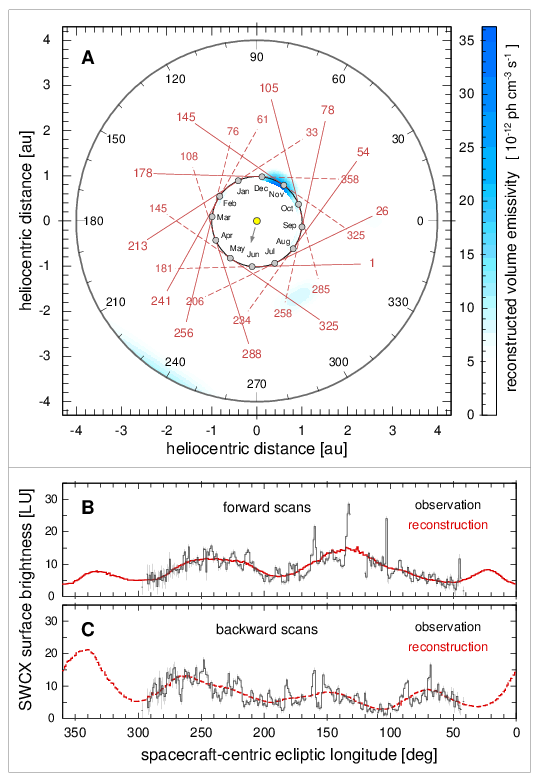}}
\caption{\textbf{Localization of the emission regions.}
{\bf (A)} Reconstructed emissivity map in the ecliptic plane, centered at the Sun (yellow circle), which is moving relative to the LISM in the direction indicated by the arrow. Gray circles indicate the position of Earth at the middle of each month. Red lines indicate the eROSITA observing directions -- solid lines are scans along the orbital motion (forward scans) and dashed lines are scans in the opposite direction (backward scans). Numbers labelled on these lines (and on the outer circle) indicate the ecliptic longitudes. {\bf (B)} Observed (black histograms with $1\sigma$ error bars) and reconstructed (red line) surface brightness profiles in the forward scans, averaged over $\pm20\deg$ in ecliptic latitude. These data are from \erass3 and \erass4; the data gaps are when eROSITA was observing outside the western Galactic hemisphere. {\bf(C)}~Same as panel (B), but for the backward scans.}
\label{fig:loc_result}
\end{figure}


\begin{figure}
\centerline{\includegraphics[height=1.0\textwidth,angle=-90]{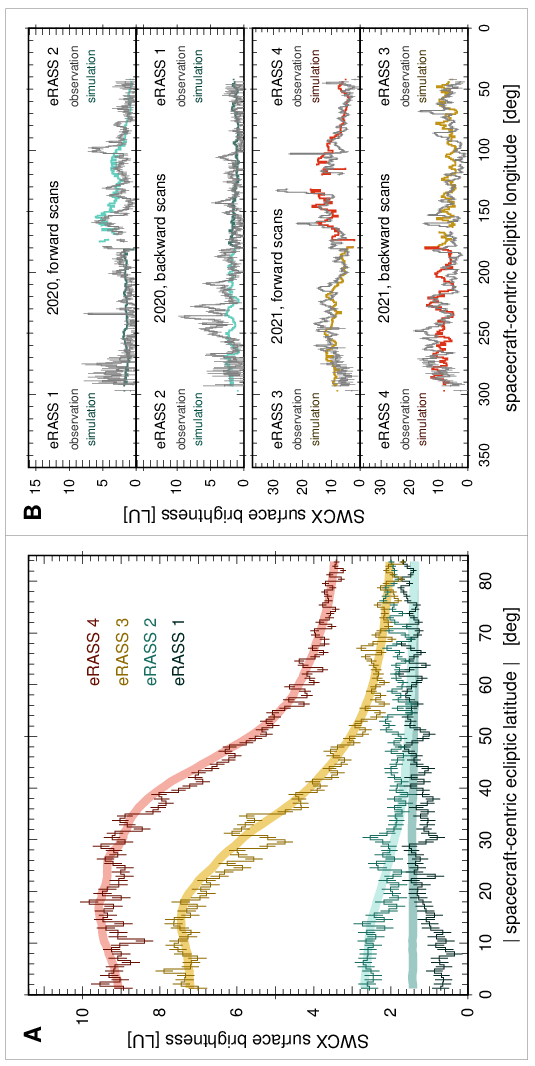}}
\vspace*{4mm}
\caption{\textbf{Heliospheric soft X--ray (0.27\,--\,0.69~keV) surface brightness profiles}.
{\bf(A)} Observed emission (histograms with $1\sigma$ error bars) as a function of spacecraft--centric ecliptic latitude for \erass1 to \erass4, compared to our simulations (thick lines). To improve the statistics, negative and positive latitudes were combined. {\bf (B)} The same data and simulations as in (A), but as functions of longitude. Forward and backward scans for pairs of surveys are shown in each subpanel, with different axis scaling.}
\label{fig:profiles}
\end{figure}


\begin{figure}
\vspace*{-8mm}
\centerline{\includegraphics[width=0.75\textwidth]{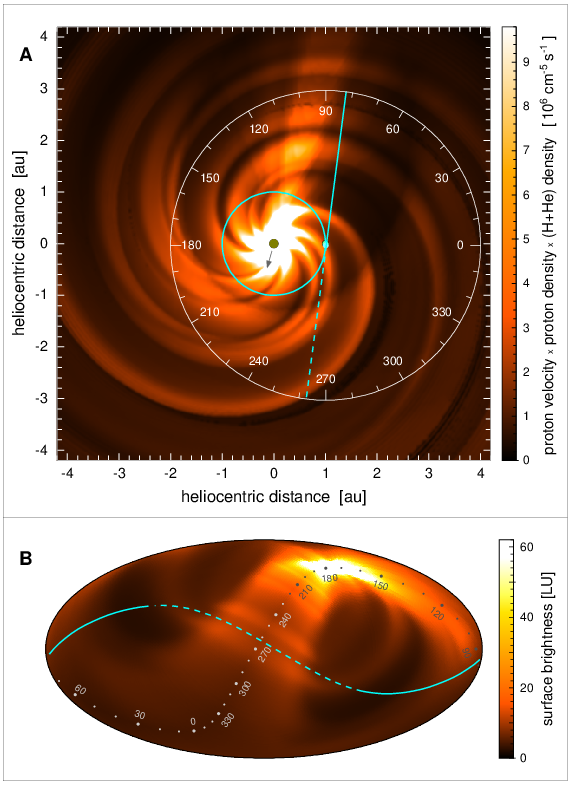}}
\caption{\textbf{Simulation of the heliospheric emission.}
{\bf (A)} Product of the solar wind proton flux and the LISM density, shown in the ecliptic plane for a specific day (21 September 2021) in our simulation. For a fixed SWCX efficiency, this product is proportional to the volume emissivity. The arrow indicates the movement of the Sun (gray circle) through the LISM. The cyan dot marks the location of Earth on its orbit (cyan circle). Cyan lines indicate the eROSITA lines of sight along the forward (solid) and backward (dashed) directions. Numbers on the outer white circle indicate ecliptic longitudes. {\bf (B)} The corresponding simulated SWCX surface brightness in the eROSITA energy band across the full sky, displayed in an equal-area Hammer-Aitoff projection. The Galactic Center is in the middle, and the Galactic plane runs horizontally. The cyan line shows the great circle observed by eROSITA on that day in the forward (solid) and backward (dashed) scan. Numbered gray dots indicate ecliptic longitudes. Movie S3 shows an animated version of this figure covering other dates.}
\label{fig:simulation}
\end{figure}


\clearpage 


\section*{Acknowledgments}
We thank the anonymous reviewers for their thorough feedback, which improved the paper substantially. This work is based on data from eROSITA, the soft X--ray instrument aboard SRG, a joint Russian--German science mission supported by the Russian Space Agency (Roskosmos), in the interests of the Russian Academy of Sciences represented by its Space Research Institute (IKI), and the Deutsches Zentrum f\"ur Luft- und Raumfahrt (DLR). The SRG spacecraft was built by Lavochkin Association (NPOL) and its subcontractors, and is operated by NPOL with support from the Max Planck Institute for Extraterrestrial Physics (MPE). The development and construction of the eROSITA X--ray instrument was led by MPE, with contributions from the Dr.~Karl Remeis Observatory Bamberg \& ECAP (FAU Erlangen--N\"urnberg), the University of Hamburg Observatory, the Leibniz Institute for Astrophysics Potsdam (AIP), and the Institute for Astronomy and Astrophysics of the University of T\"ubingen, with the support of DLR and the Max Planck Society. The Argelander Institute for Astronomy of the University of Bonn and the Ludwig Maximilians Universit\"at Munich also participated in the science preparation for eROSITA. The eROSITA data were processed using the eSASS software system developed by the German eROSITA consortium.

\paragraph*{Funding:}
G.~P.~was supported by the European Research Council (ERC) under the European Union's Horizon 2020 research and innovation program Hot-Milk (grant agreement No 865637) and from the Framework per l'Attrazione e il Rafforzamento delle Eccellenze (FARE) per la ricerca in Italia (grant number R20L5S39T9). 
M.~C.~H.~Y.~and M.~J.~F.~were supported by the Deutsche Forschungsgemeinschaft through the grant FR 1691/2-1.

\paragraph*{Author contributions:}
K.\,D. performed the data analysis and led the study and manuscript preparation.
G.\,P. intiated the study, identified SWCX in the eROSITA data, and mapped the ecliptic enhancements.
X.\,Z. produced the eRASS maps.
M.\,J.\,F. investigated the particle background.
S.\,F. and Th.\,M. contributed to the discussion and manuscript preparation.
M.\,C.\,H.\,Y. contributed to the data analysis and manuscript revision.

\paragraph*{Competing interests:}
G.~P. is also affiliated with the Como Lake Center for Astrophysics, Università degli Studi dell’Insubria, Como, Italy. The authors declare no other competing interests.

\paragraph*{Data, code, and materials availability:}

The eROSITA data used in this study (smoothed 0.27\,--\,0.69 keV eRASS maps) and our derived dark sky map are archived at Zenodo \cite{zenodo_doi}. Our data analysis software and simulation software are also available in the same Zenodo repository \cite{zenodo_doi}. The solar wind data were obtained from the OMNIWeb database \cite{omniweb}. No physical materials were generated in this work.

\newpage


\renewcommand{\thefigure}{S\arabic{figure}}
\renewcommand{\thetable}{S\arabic{table}}
\renewcommand{\theequation}{S\arabic{equation}}
\renewcommand{\thepage}{S\arabic{page}}
\setcounter{figure}{0}
\setcounter{table}{0}
\setcounter{equation}{0}
\setcounter{page}{1} 


\begin{center}
\section*{Supplementary Materials for\\ \quad\scititle}

K. Dennerl$^{\ast}$,
G. Ponti,
X. Zheng,
M. J. Freyberg,
S. Friedrich,
Th. M{\"u}ller,
M. C. H. Yeung\\
\small$^\ast$Corresponding author. Email: kod@mpe.mpg.de\\

\end{center}

\subsubsection*{This PDF file includes:}
Materials and Methods\\
Supplementary Text\\
Figures S1 to S5\\
Captions for Movies S1 to S5\\

\subsubsection*{Other Supplementary Materials for this manuscript:}
Movie S1 to S5


\newpage

\subsection*{Materials and Methods}

\subsubsection*{eROSITA flux measurements}

eROSITA is equipped with seven essentially identical coaligned Telescope Modules (TMs), abbreviated with TM1\,--\,TM7 \cite{pre21a}. After launch it turned out that two of them, TM5 and TM7, are affected by sunlight which reaches the cameras sideways. This causes the soft X--ray signal to be severely contaminated. In order to avoid possible misinterpretations, we base our analysis on the data obtained with TM\,1,2,3,4,6. This combination is usually referred to as TM\,8. The charge released by the absorption of a photon in a CCD is not necessarily constrained to one pixel, but may extend over a pattern consisting of several of them, in the case of eROSITA of up to four \cite{den20a}. In order to maximize the sensitivity, we utilize all these patterns. When dealing with diffuse emission, the photon flux is often expressed in `line units' (LU), defined as photons per square centimeter per second per sterad (photons cm$^{-2}$ s$^{-1}$ sr$^{-1}$) in a given energy band. For comparison with previous studies we express the eROSITA flux in LU.

\subsubsection*{Extracting the heliospheric emission component}

Our analysis is based on four eROSITA sky maps, obtained during \erass1 to \erass4, in the energy range 0.269 to 0.694 keV (Supplementary Text). They were generated with version c020 of the eROSITA Science Analysis Software System pipeline (eSASS, \cite{bru22a}) and contain the Western Galactic hemisphere in an ecliptic equirectangular projection with a pixel size of $5\mbox{ arcmin}\times5\mbox{ arcmin}$ at the ecliptic plane.

In a first step, we reduce high-frequency noise and the impact of bright, potentially variable point sources on the diffuse emission. For each map, we determine the flux value exceeded by the brightest 0.5\% of pixels and store the positions of these pixels in a separate mask. The flux maps are then smoothed by replacing each pixel by the average of its $5\times5$ neighborhood, ignoring masked pixels; they are set to the previously determined flux threshold \cite{code:prep1}. We then subtract the instrumental background \cite{code:prep2}, derived by scaling the flux in 5 to 10~keV maps (which are dominated by instrumental background) to our energy range, using the spectral information obtained from measurements with the filter wheel closed (\cite{yeu23a}, their appendix A). The resulting maps (A$_1$ to A$_4$) are shown in Fig.\,\ref{fig:heliosphere}A-D.

For composing the dark sky map, we need to determine for each celestial region the survey scan number (1 to 4 for \erass1 to \erass4) which yields the minimum flux. Here we need to consider that, due to photon statistics, taking the minimum value of four measurements may result in a map which contains too little flux. Even if the same flux was measured four times, then the dark sky map would contain (slighty) less flux than the individual input maps. This is an inherent problem which, given the stochastic nature of the X--ray flux, cannot be avoided. However, we attempt to minimize its effect by ensuring that the photon statistics in all sky regions is sufficiently high.

The statistical fluctuations had already been reduced by smoothing the non--masked areas of the maps with a running mean computed over a $5\times5\mbox{ pixel}$ wide kernel. Now we compute a temporary version of those A$_1$ to A$_4$ maps where we apply an additional substantial smoothing along the scan direction, utilizing the fact that it is unlikely that heliospheric enhancements occur on time scales shorter than $\sim10\mbox{ min}$: we smooth with a running average over $\pm\,60\mbox{ pixels}$ or $\pm\,5\deg$ in ecliptic latitude (again ignoring the masked pixels). At a scanning speed of $1.5\deg\mbox{ min}^{-1}$, this corresponds to $\pm\,200\mbox{ s}$. We use these maps to determine an eRASS selection map, which contains for each pixel the eRASS number for which the 0.27 to 0.69~keV flux in that pixel is lowest. This eRASS selection map is then used to compose the dark sky map by taking for each pixel the flux from the corresponding A$_1$ to A$_4$ maps \cite{code:prep3}.

We then subtract the dark sky map from the eRASS maps A$_1$ to A$_4$, to produce maps of the excess emission (Fig.\,\ref{fig:heliosphere}\,F-H). These maps contain episodes of enhanced particle background. We exclude them with mask maps created from the 5 to 10~keV maps, which are dominated by the particle background, and get maps B$_1$ to B$_4$, which are displayed in Fig.\,\ref{fig:heliosphere}\,I-L after having applied a running median filter with a $\pm4\deg$ kernel along the scanning direction to enhance the SWCX emission; this filtering was only done for display purposes. The next step is to compute the pixel-dependent exposure time for each of the four eRASS maps as well as for the dark sky map \cite{code:prep6}. This is necessary for the determination of the $1\sigma$ errors. Then the surface brightness profiles in Figs.\,\ref{fig:loc_result} and {\ref{fig:profiles}} can be computed from the maps B$_1$ to B$_4$ \cite{code:flxobs}.

\subsubsection*{Localization of the emitting regions}

In order to apply the method outlined in Fig.\,\ref{fig:loc_demo} in an unbiased way, we fit an emissivity map to the observed surface brightness profiles. We do this by scanning a candidate map in the same way as eROSITA had scanned the sky, computing the resulting profiles and repeating this process until the derived profiles match the observed ones.

One requirement for applying this method is to find a suitable parameterization of an unbiased emissivity map, because this requires a large number of free parameters. A map with $100\times100$~pixels would contain $10^4$~pixels, and each pixel would be a free parameter. To reduce the number of free parameters, we distribute a total of 72 emission centers in the ecliptic plane over a $8.6\times8.6\mbox{ au}$ wide area around the Sun  (Fig.\,\ref{fig:loc_par}A). We do not place them inside Earth's orbit, because these regions were not observed in the eROSITA scans. These emission centers are used as the knots of a bicubic spline interpolation, performed on a Cartesian grid (Fig.\,\ref{fig:loc_par}B) to preserve azimuthal structures. We replace negative values in the interpolation (which would correspond to absorption) by zeros.

The next requirement is to find a fast method for computing surface brightness profiles, because this requires to sample in each iteration the emissivity map in the same way as eROSITA had scanned the sky, and several $10^5$~iterations are necessary to determine the 72~free parameters. To reduce the computational cost, we pre-calculate for each time step the set of pixels in the emissivity map which contribute to the observed flux and store their positions. This avoids the necessity to repeat geometrical calculations in every time step. We then pre-compute the contribution of each of the 72~emission centers to the surface brightness profiles \cite{code:loc01}, then search for a linear combination of these basis functions that reproduces the observed profiles, while requiring that no coefficient is negative. This is achieved by expressing the scaling factor of each curve as the square of the actual fit parameter \cite{code:loc02}.

We extract the flux profiles from a region that covers the ecliptic latitudes $\pm\,20\deg$, but consider only ecliptic longitudes which cover at least 10\% of the $\pm\,20\deg$ full latitudinal extent (the reduced coverage in ecliptic latitude is a consequence of the restriction to the Western Galactic hemisphere). We focus on the \erass3 and \erass4 data, because they provide the best signal.

The emissivity map is computed with a resolution of $512\times512$ pixels. Surface brightness profiles are derived by extracting the emissivity values along the line of sight up to a distance of 4.3~au (the maximum distance which is fully covered by the emissivity map along the scanning directions), with an opening angle of $1\deg$, corresponding to the eROSITA field of view. This was done by sampling the emissivity map with a radial step size of 1~pixel ($2.5\times10^6\mbox{ km}$) and an azimuthal step size of $0.1\deg\!$, averaging the emission azimuthally for each radius, then summing up these values. Despite the complexity of this algorithm, it runs very fast: convergence is achieved after $\sim400\,000$~iterations, which take $\sim40\mbox{ s}$ on an Apple M3 processor \cite{code:prloc}.

\subsubsection*{Comparison to simulations}

The input solar wind measurements \cite{omniwebdata} were obtained from near the $L_1$ point of the Sun--Earth system. To reconstruct a map of the proton flux over the inner Solar System, we compute its propagation under the assumption that the measured flux originates from localized active regions on the Sun which persist for one solar rotation. This is a simplifiying assumption, but the coronal configuration near solar minimum can be stable over many months [\cite{bal09a}, their figure 7~]. There are, however, also cases where such an assumption is not applicable, particularly coronal mass ejections; these are not covered by our simulation.

By tracing the solar wind flows measured near $L_1$ back to the Sun, we determine their footprints and construct a solar activity profile, containing the proton densities and velocities as a function of solar longitude. To compute a map of the proton flux over the ecliptic plane at time $t_0$, we use the in--situ measurements at $L_1$ taken between $t_0-20\mbox{ d}$ and $t_0+12\mbox{ d}$ for the solar activity profile. We then sample the profile in steps of $0.5\deg\!$, compute for each solar longitude the path of a solar wind parcel which was ejected radially from that longitude over a time period $\left[t_0-30\mbox{ d}\mbox{ to }t_0\right]$ with a step size of 0.1~d, then populate an initially empty $512\times512\mbox{ pixel}$ map of $8.6\times8.6\mbox{ au}$ around the Sun with the proton flux in that parcel. In order to avoid sharp edges and empty regions in the proton flux map, we vary the solar wind velocity by $\pm20\mbox{ km s}^{-1}$ around the measured value in 7 equidistant steps, compute for each velocity the path, and distribute the proton flux over these paths by a weighted average, with factors of 0.05, 0.1, 0.2, 0.3, 0.2, 0.1, 0.05, which approximates a Gaussian distribution with $\sigma\sim8\mbox{ km s}^{-1}$. To check the results, we compare the density and velocity maps computed in this way with those obtained from calculations by the coupled solar wind (WSA) and heliospheric propagation (ENLIL) model used for CME and solar wind forecasting \cite{ods99a,arg00a}. We find differences between the computations, but no more than we expected due to our simplified assumptions (Fig.\,\ref{fig:comparison}). We find also discrepancies in the WSA–ENLIL predictions at identical time instants when the two simulations are initiated from different starting dates. Previous comparisons of WSA–ENLIL predictions of corotating interaction regions with in situ measurements indicate typical timing errors of $\sim2\mbox{ days}$ at 1 au and $\sim4\mbox{ days}$ at 5.4 au \cite{lee09a,jia11a,pri15a}.

For the distribution of the LISM, we assume that the densities of interstellar H and He shown in [\cite{rin21a}, their figure 1 ] are rotationally symmetric with respect to an axis pointing to longitude $74.7\deg$, latitude $-5.3\deg$ \cite{moe04a} and are applicable to the time period from \erass1 to \erass4. The product of proton flux and LISM density yields  the collision rate density per unit cross section (Fig.\,\ref{fig:simulation}A). This quantity is, for a constant SWCX efficiency, proportional to the volume emissivity. By integrating the volume emissivities along the eROSITA line of sights, we obtain a map with the SWCX surface brightness over the full eROSITA sky (Fig.\,\ref{fig:simulation}B). By comparing those parts of the map which were scanned by eROSITA with the measurements, we empirically determine the initially unkown SWCX efficiency as a function of latitude and time (Fig.\,\ref{fig:pleff}). The dip at $<10\deg$ in \erass3 and \erass4 is probably an artefact caused by residual flux in the dark sky map (Fig.\,\ref{fig:profiles}\,A). As a by-product of our simulation we obtain a reconstruction of the 0.3-0.7~keV SWCX emission over the full sky and over two years \cite{code:runsim}, from which we created Movies S3 and S4.

%
\subsection*{Supplementary Text}

\subsubsection*{SWCX emission in the 0.267 to 0.694 keV energy band}

We selected this energy range because it contains the bulk of the SWCX flux from carbon, nitrogen, and oxygen ions, dominated by C~\textsc{v}, N~\textsc{vi}, and O~\textsc{vii} emission. After electron capture, all of these ions are in a He-like state and de-excite by emitting a multiplet with centroids near 0.302, 0.425, and 0.567~keV. We have increased the upper energy boundary to include also emission from O~\textsc{viii} at 0.654~keV, and have then widened the energy band by $-35\mbox{ eV}$ and $+40\mbox{ eV}$ to take the instrumental energy resolution [\cite{den20a}, their table 1] into account.

\subsubsection*{Spatial resolution of the eROSITA maps}

eROSITA has mapped the sky with an average spatial resolution of $\sim30''$ half energy width \cite{mer24a}. However, for studying heliospheric emission, the high resolution cannot be utilized, because the emission is too faint. On the other hand, the emission is also so diffuse that high spatial resolution is not needed. We reduced the spatial resolution of the eROSITA maps, but only along the scanning direction, to preserve temporal SWCX variations down to $\sim10$~minutes. The smoothing was done by a random permutation of the pixels along the scanning direction: along each 5~arcmin wide scan path, segments of 25 consecutive pixels (stretching over 125~arcmin) were 2000 times randomly selected, and in each segment the pixels were rearranged by a random permutation. This caused structures from beyond the heliosphere to get smoothed out, while leaving heliospheric emission intact. Because no pixel values are altered (they are just rearranged along the scan direction) and because the point source suppression and masking is done on individual pixel values, these methods are also applicable to the modified maps.

\subsubsection*{Excluding residual emission in the dark sky map caused by variability on time scales $<2\mbox{ yr}$ }

As each region of the sky was observed every 4 hours, with the scan axis slowly drifting in ecliptic longitude, variation of the heliospheric emission would have caused flux variations between individual scans, which would have propagated as a longitudinal modulation into the dark sky map.

\subsubsection*{Impact of the O$^{7+}$/O$^{6+}$ ratio on the observed SWCX emission}

After charge exchange, O$^{7+}$ and O$^{6+}$ solar wind ions become O$^{6+}$ and O$^{5+}$ ions, generating O~\textsc{vii} and O~\textsc{vi} emission. While the O~\textsc{vii} emission is covered by our energy band, this is not the case for the O~\textsc{vi} emission, which is below 0.20~keV.

\subsubsection*{Quantifying the residual SWCX emission in the dark sky map}

This can be deduced directly from the latitudinal SWCX surface brightness profiles obtained in \erass1 (Fig.\,\ref{fig:profiles}A): the mean value for latitudes $|l|\le17.5\deg$ ($0.83\pm0.05\mbox{ LU}$) is $0.59\pm0.05\mbox{ LU}$ below the mean value for $|l|\ge20.0\deg$ ($1.42\pm0.017\mbox{ LU}$). Taking potential systematic uncertainties into account, we quote the residual SWCX emission as $\sim1\mbox{ LU}$.

\subsubsection*{eROSITA sensitivity limit}

Each of the values in the latitudinal \erass1 SWCX surface brightness profiles  (Fig.\,\ref{fig:profiles}A) was obtained over a latitudinal range of $2\times50\mbox{ arcmin}$ (2 because positive and negative latitudes were combined) and a longitudinal range of $180\deg\times\cos\left(l\right)$ ($180\deg$ because of the restriction to the Western Galactic hemisphere), yielding a solid angle of $300\times\cos\left(l\right)$ square degrees. Despite this large solid angle, the mean deviation of the values below latitude $|l|\le17.5\deg$ from the average value for latitudes $|l|\ge20.0\deg$ ($1.42\pm0.017\mbox{ LU}$) is $2.7\sigma$ and thus only marginally significant.

\subsubsection*{Instantaneous and average solar wind}

While the low latitude solar wind exhibits a lot of variability, this variability gets smoothed out when averaged over more than a few days, so that the solar wind can be approximated by a constant flow (Movie S5).


\begin{figure}
\centerline{\includegraphics[width=0.8\textwidth]{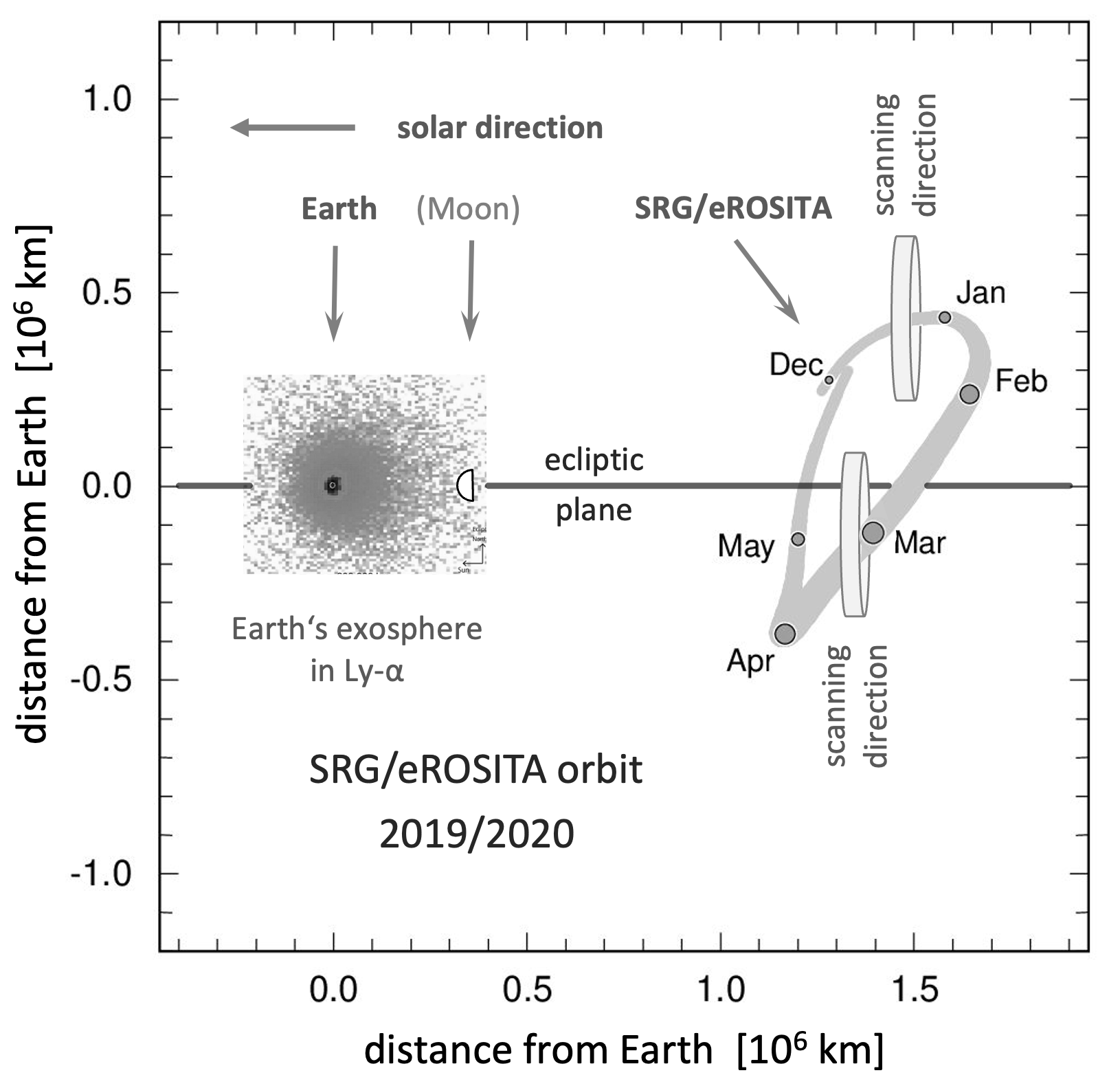}}
\smallskip
\caption{\textbf{Orbit of SRG/eROSITA and observing geometry.}
Projection of the SRG/eROSITA orbit during \erass1 onto a plane perpendicular to the Earth orbit and passing through the Sun and Earth, thus corotating with the Earth around the Sun (indicated by the solar direction arrow). Filled circles mark the position of SRG/eROSITA at the middle of each month; their size indicates the distance from the projection plane. A Ly-${\alpha}$ image of Earth \cite{kam17a} is inserted to scale, together with a Moon symbol, to indicate the extent of the geocorona. The two cylindrical surfaces on the SRG/eROSITA orbit visualize the great circles along which the sky was scanned, and demonstrate that the eROSITA view was not affected by geocoronal emission. The Ly-${\alpha}$ image, adapted from [\cite{kam17a}, their figure 1a] was reproduced with permission.}
\label{fig:orbit}
\end{figure}


\begin{figure}
\vspace*{-1cm}
\centerline{\includegraphics[width=0.85\textwidth]{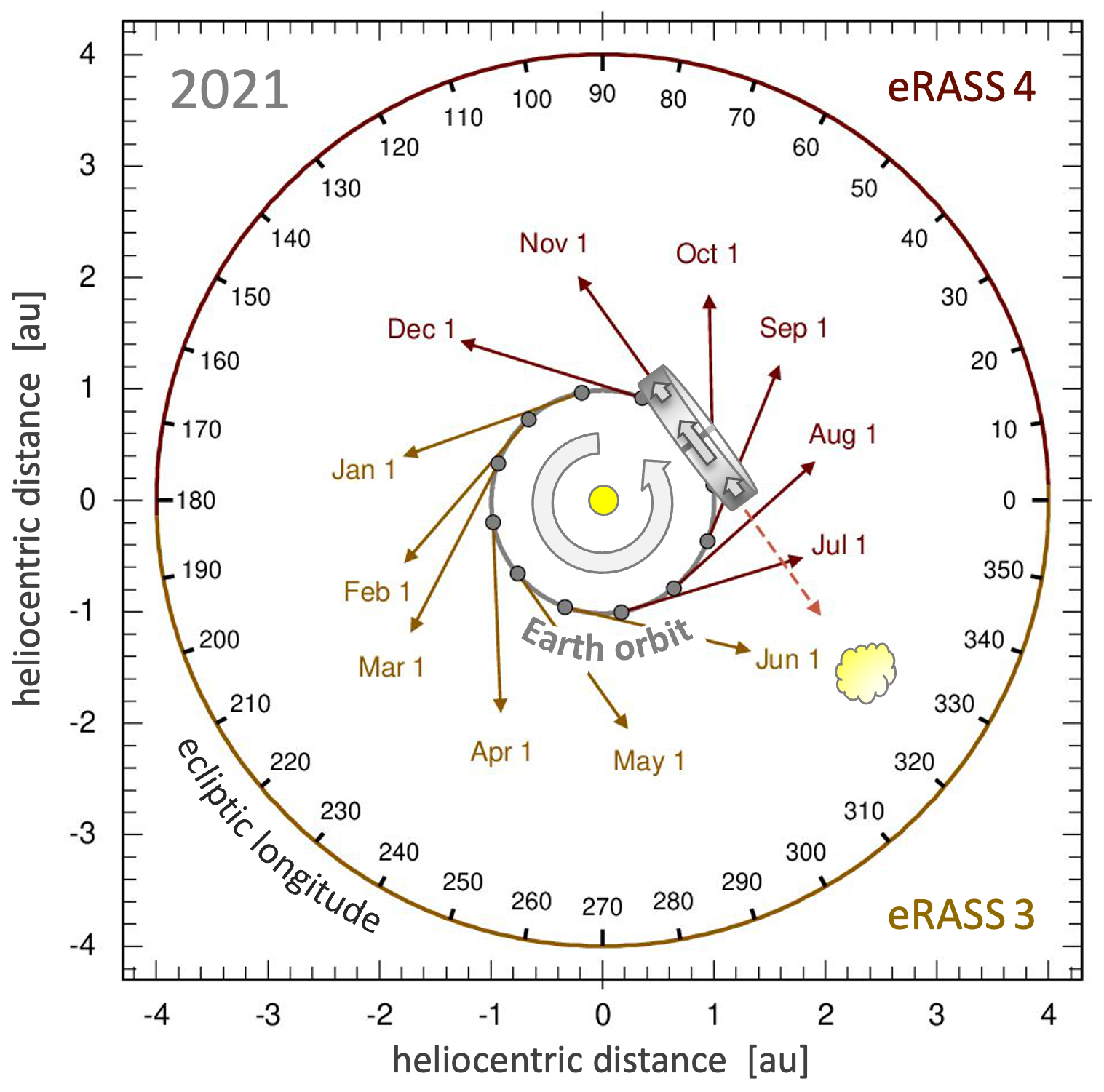}}
\caption{\textbf{Illustration of the method of localizing the emission regions.} View of the Earth orbit in the ecliptic plane (counterclockwise, indicated by the large circular arrow), with the Sun at center (yellow circle) and the location of Earth plotted for every month. A cylindrical surface with arrows visualizes the great circle along which the sky was scanned on Nov~1. The scan axis was permanently kept in the ecliptic plane, pointing toward the Sun within $13^{\circ}$. Arrows, drawn for the beginning of every month, show which ecliptic longitudes (specified by the numbers on the large circle) were scanned at those dates along the direction of Earth's motion (forward scans). The arrow for June~1 points towards a stationary cloud. By using the information that this cloud was additionally only observed in the backward scan of Nov~1 (dashed line), this cloud can be localized. As the compilation of one full-sky map takes only half a year, this method requires two such maps, here illustrated for \erass3 (brown) and \erass4 (red).}
\label{fig:loc_demo}
\end{figure}


\begin{figure}
\vspace*{-1cm}
\centerline{\includegraphics[width=0.75\textwidth]{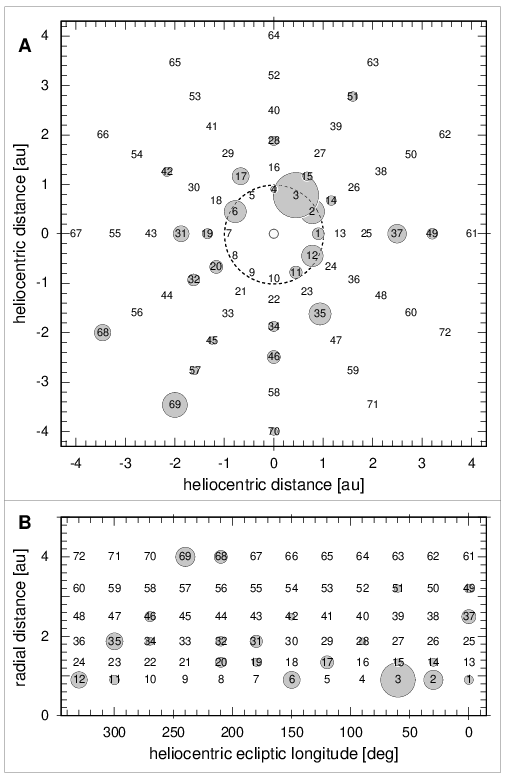}}
\caption{\textbf{Results obtained with the method of localizing the emission regions.}
{\bf A:} View of the Earth orbit (dashed) around the Sun (central small circle) in the ecliptic plane, with the location of the 72 emission regions marked by running numbers. The areas of the filled gray circles indicate the emissivities found. {\bf B:} Same as {\bf A}, but plotted in the coordinate system which was used for the bicubic spline interpolation.}
\label{fig:loc_par}
\end{figure}


\begin{figure}
\includegraphics[width=1.0\textwidth]{comparison}
\caption{\textbf{Comparison of the plasma density and radial velocity calculations.}
{\bf A, B:} Results obtained for 2021~Jan~01 from the Wang–Sheeley–Arge ENLIL (WSA-ENLIL) model calculations \cite{ods99a,arg00a}, performed at the National Aeronautics and Space Administration Community Coordinated Modeling Center (NASA CCMC).
{\bf C, D:} Results obtained for the same date with the method used for our simulation. Note that in {\bf A} and {\bf C} the densities were multiplied with the square of the heliocentric distance.}
\label{fig:comparison}
\end{figure}


\begin{figure*}[ht]
\centerline{\includegraphics[width=\textwidth]{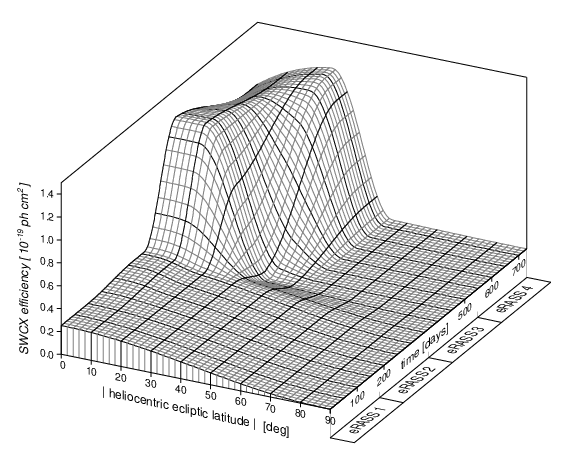}}
\caption{\textbf{SWCX efficiencies derived from the simulation.}
The curves show the scaling factors used to convert the product of solar wind flux and LISM density in the ecliptic plane into 0.27 to 0.69~keV volume emissivities. They were derived for all heliocentric ecliptic latitudes and the full \erass1 to \erass4 period. Integration of the resulting volume emissivities along the line of sight yields the SWCX fluxes in Fig.\,\ref{fig:profiles}.
}
\label{fig:pleff}
\end{figure*}


\clearpage 


\begin{figure}
\centerline{\includegraphics[width=\textwidth]{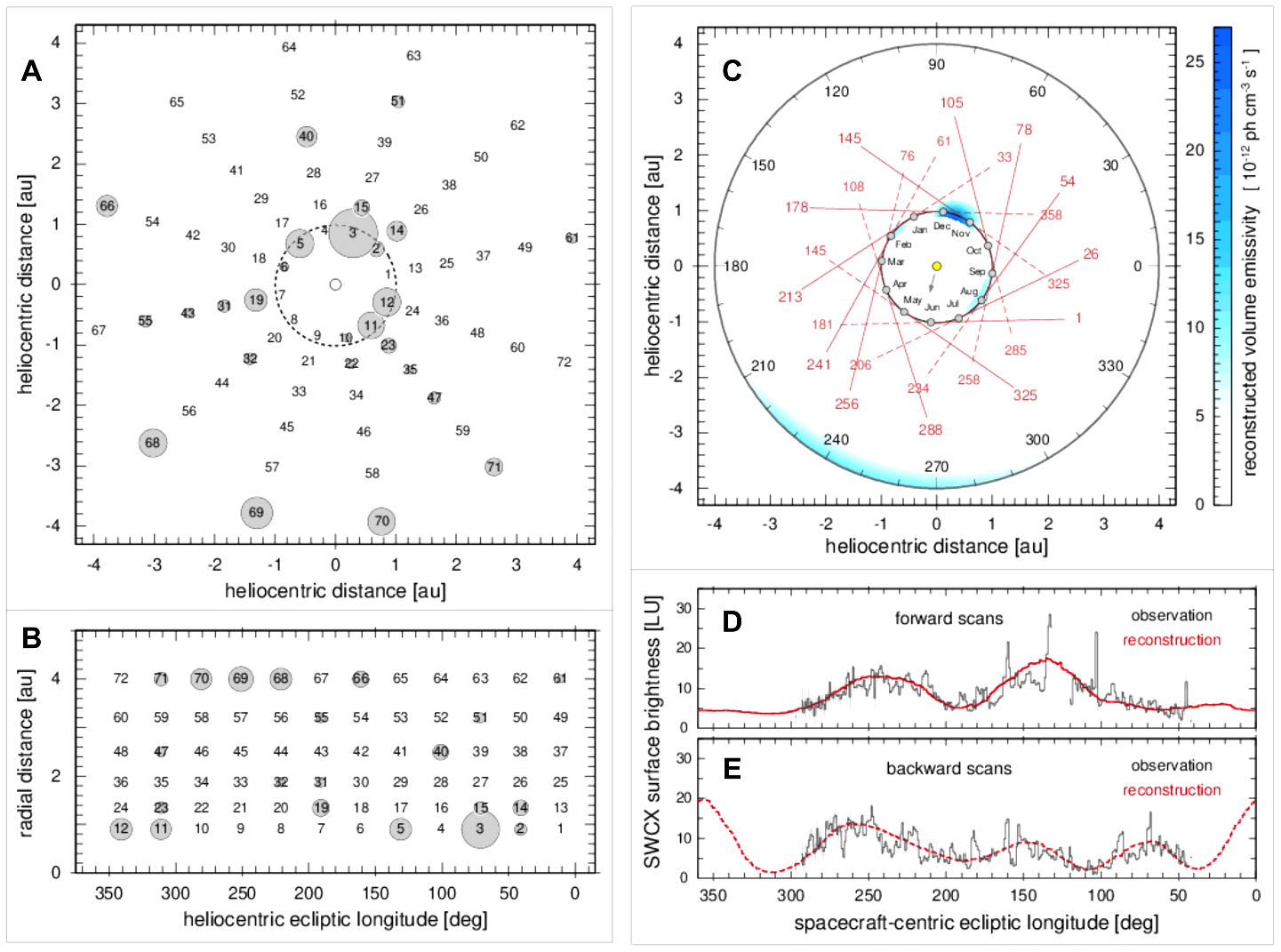}}
\bigskip
{\textbf{Movie S1: Localization of the heliospheric X--ray emission.}
{\bf A,B:} Same as Fig.\,\ref{fig:loc_par}\,A,B, {\bf C-E:} same as Fig.\,\ref{fig:loc_result}\,A-C, but for different placements of the emission centers, which are rotated in steps of $1\deg$ counterclockwise around the Sun. This movie is best displayed in a loop.}
\label{movie_s1}
\end{figure}


\begin{figure}
\centerline{\includegraphics[width=\textwidth]{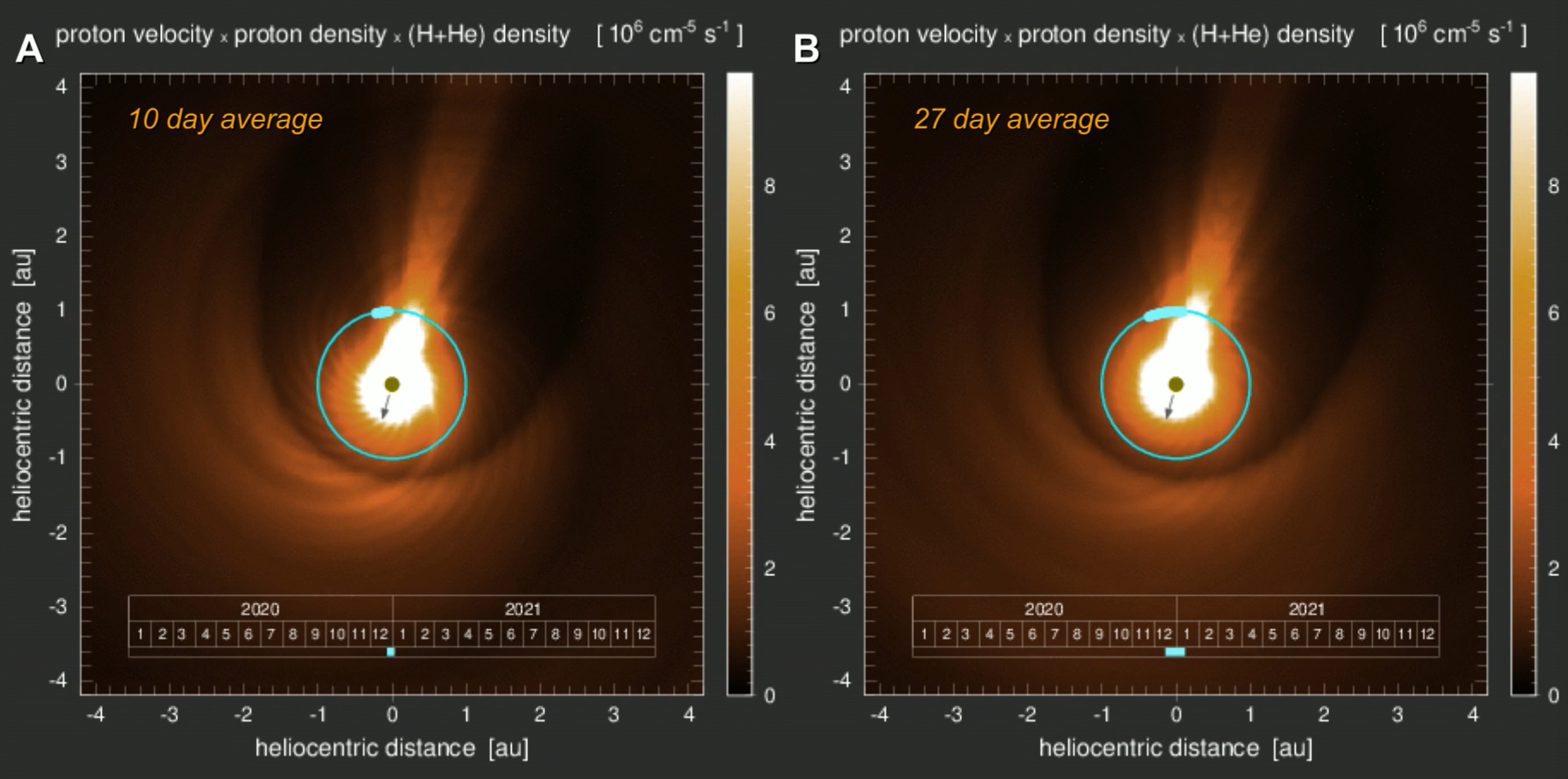}}
\bigskip
{\textbf{Movie S2: Stability of the He focusing cone.}\
\smallskip
Maps of the product of the solar wind proton flux with the LISM density in the ecliptic plane, centered at the Sun (grey circle; its movement through the LISM is indicated by an arrow). The blue circle indicates the Earth orbit, where the location of SRG/eROSITA is shown for the time interval which is marked by the blue line in the calendar below. This time interval is 10 days in {\bf A} and 27 days (about one solar rotation) in {\bf B}. The inhomogeneities of the solar wind cause the SWCX emission of the He focusing cone to be variable, even in 10~day averages ({\bf A}). By averaging over one solar rotation (right), the variability gets considerably suppressed ({\bf B}).}
\label{movie_s2}
\end{figure}


\begin{figure}
\vspace*{-2cm}
\centerline{\includegraphics[width=\textwidth]{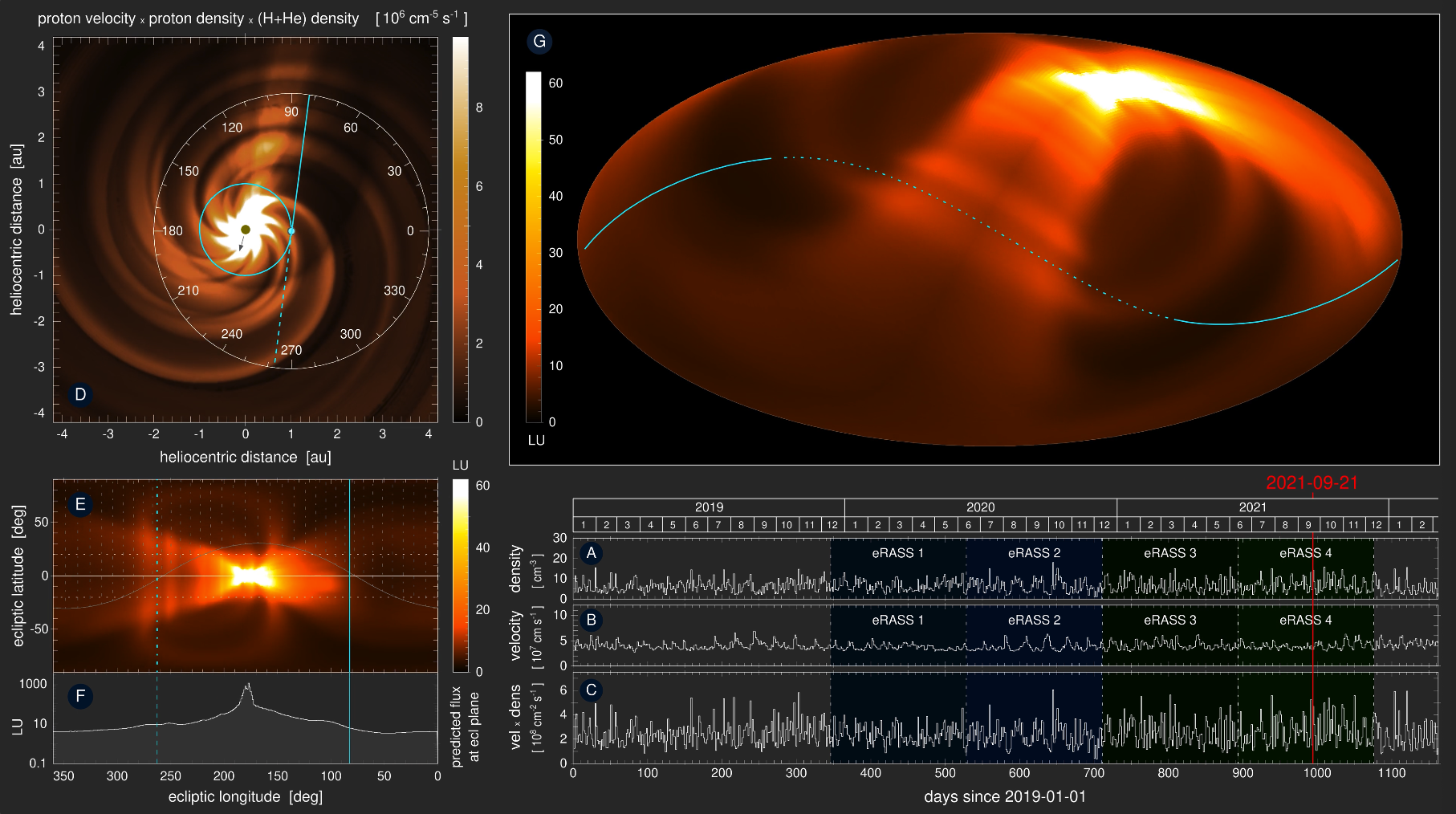}}
\bigskip
{\textbf{Movie S3: Animated simulation}.}
{\bf A:} Measured solar wind proton density, {\bf B:} measured solar wind velocity, {\bf C:} product of velocity and density, yielding the proton flux. {\bf D:} Map with the product of proton flux and LISM density in the ecliptic plane, centered at the Sun (grey circle; its movement through the LISM is indicated by an arrow). The blue circle indicates the Earth orbit, where the location of SRG/eROSITA is shown for the date marked in red in {\bf A-C}. Blue straight lines show the observing direction in the forward (solid) and backward (dashed) scans. Numbers along the large white circle indicate ecliptic longitudes. This map is essentially the same as that in Fig.\,\ref{fig:simulation}\,A. {\bf E:}~Resulting SWCX surface brightness in the eROSITA sky in an equirectangular ecliptic projection, with the scanning directions marked by blue lines like in {\bf D}. The dashed horizontal lines indicate the latitude range which was used for extracting the longitudinal profiles in Figs.\,\ref{fig:loc_result}\,B,C and \ref{fig:profiles}\,B. The sinusoidal white curve indicates the boundary of the Western Galactic hemisphere, which does not include the regions marked by vertical dashed white lines. {\bf F:} Profile of the SWCX surface brightness seen along the ecliptic plane. {\bf G:} Same as {\bf E}, but displayed in an equal area Hammer--Aitoff projection, with the Galactic center in the middle and the Galactic plane running horizontally. This map is essentially the same as that in Fig.\,\ref{fig:simulation}\,B.
\label{movie_s3}
\end{figure}

\clearpage

\begin{figure}
\centerline{\includegraphics[width=\textwidth]{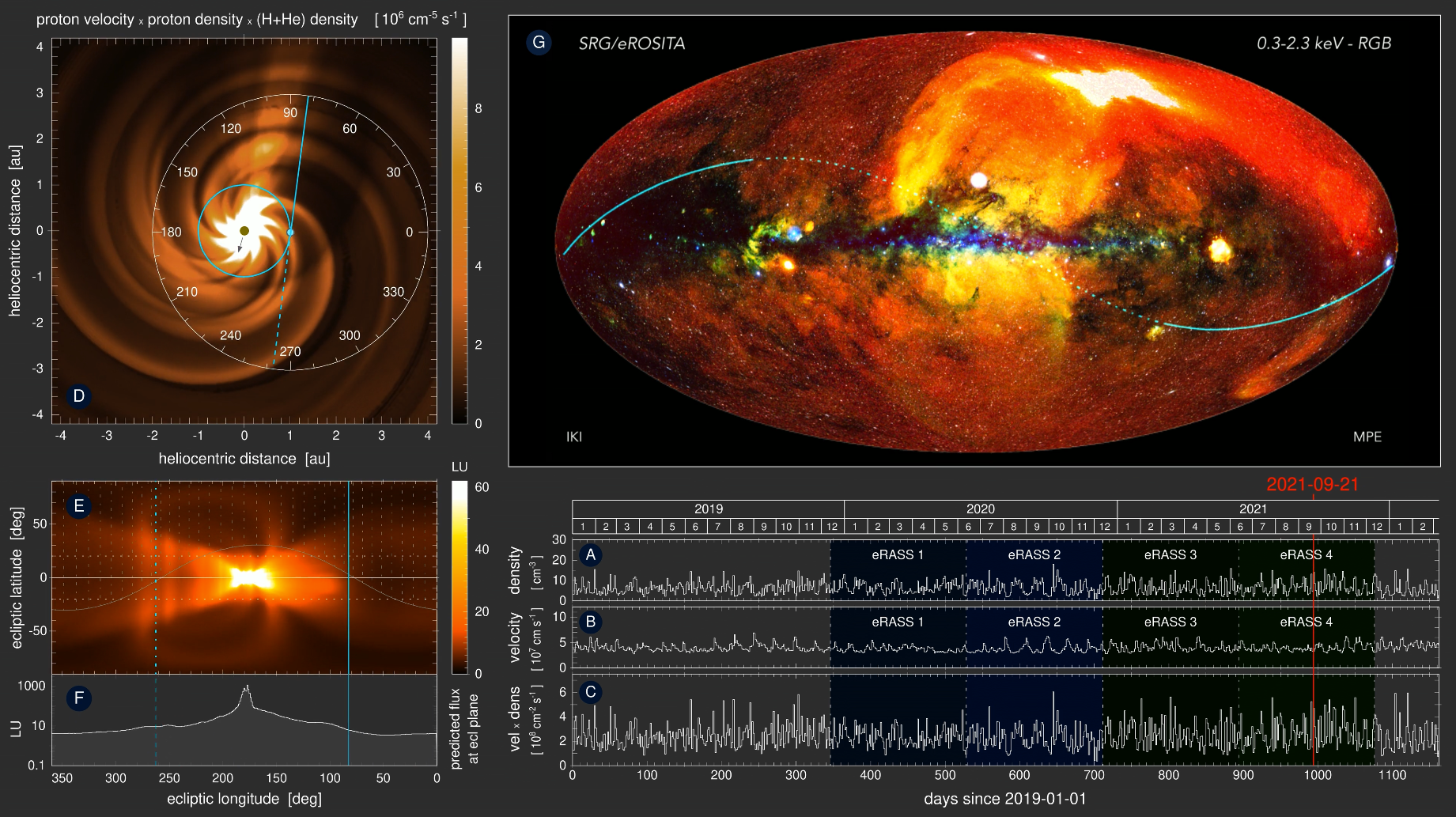}}
\bigskip
{\textbf{Movie S4: Animated simulation with eROSITA sky.} 
This movie is identical to movie S3, except that the SWCX surface brightness map in {\bf G} was superimposed on an eROSITA all-sky RGB image in approximately the correct color and intensitiy, to reconstruct the eROSITA sky during the \erass1 to \erass4 period. eROSITA, however, has only observed the sky along the blue curve.
}
\label{movie_s4}
\end{figure}


\begin{figure}
\centerline{\includegraphics[width=\textwidth]{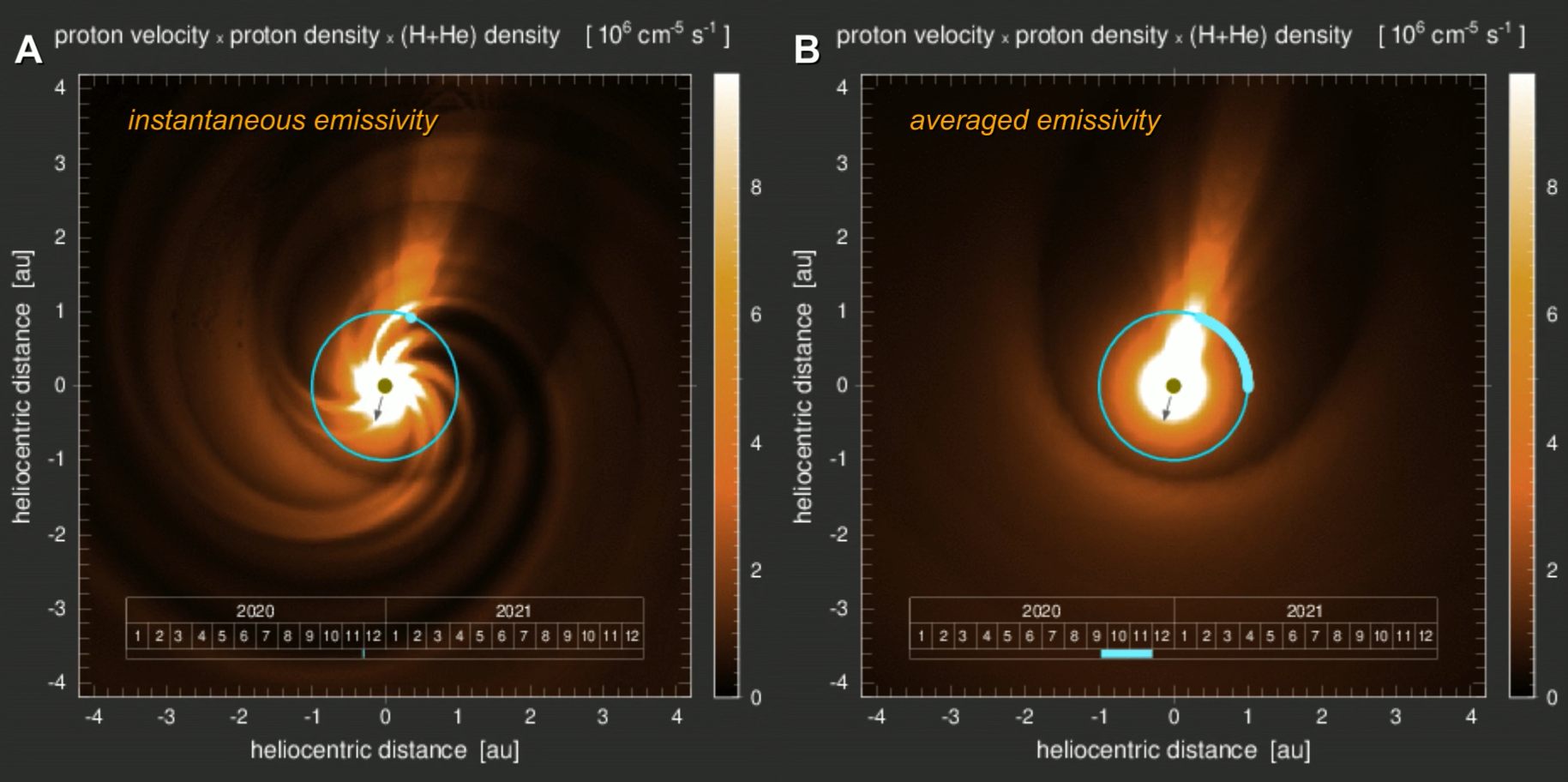}}
\bigskip
{\textbf{Movie S5: Dominant structures in the instantaneous and averaged emissivity map.}
This movie shows the same map of the collision rate density per unit cross section
as movie S2 which is, for a constant SWCX efficiency, proportional to the volume emissivity. The instantaneous view ({\bf A}) is compared with one which is averaged over an increasingly longer time, keeping the start time fixed ({\bf B}). It illustrates that the instantaneous view is dominated by spiral emissivity enhancements, which average out for longer integration times, so that the persistent LISM distribution, especially the helium focusing cone, becomes the dominant structure.}
\label{movie_s5}
\end{figure}



\begin{thebibliography}{10}
\bibitem{note:heliosphere}
For a line of sight crossing the heliosphere, the neutral hydrogen column density is $N_{\rm H}\sim10^{13}$ to $10^{15}\mbox{ cm}^{-2}$
\bibitem{note1}
The radial expansion of the solar wind causes the SWCX volume emissivity to
  decrease with the inverse square of the heliocentric distance. For emission
  confined to a thin shell, the observed surface brightness follows the same
  dependence. Accordingly, the surface brightness of SWCX emission at 3.2 au is
  only about 10\% of that at 1 au.
\bibitem{code:prep1}
These initial steps were performed using the program {\tt prep1} in \cite{zenodo_doi}.
\bibitem{code:prep3}
The dark sky map is computed using the program {\tt prep3} in \cite{zenodo_doi}.
\bibitem{code:flxobs}
This is performed using the program {\tt flxobs} in \cite{zenodo_doi}.
\bibitem{code:loc01}
The placement of the emission regions and the computation of the basis
  functions are done in the program {\tt loc01} in \cite{zenodo_doi}.
\bibitem{code:loc02}
The fitting is done in the program {\tt loc02} in \cite{zenodo_doi}.
\bibitem{code:runsim}
All-sky maps for each day from \erass1 to \erass4 can be generated with the
  program {\tt runsim} in \cite{zenodo_doi}, which also extracts the regions
  that were scanned.
  This information is used by the program {\tt mapsim} to compose all-sky maps
  that eROSITA should have seen in each eRASS. The program {\tt flxsim}
  extracts from these maps the surface brightness profiles resulting from the
  simulation, which {\tt prflx} compares with the observed ones.
\end{thebibliography}
\end{document}